\documentclass[smallextended]{svjour3}
\usepackage{graphics}
\usepackage{amsmath,amssymb}
\usepackage{natbib}
\usepackage{bm}
\usepackage{url}

\smartqed
\journalname{Exp Astron}

\def\ct{\ensuremath{c_{\rm t}}}\def\st{\ensuremath{s_{\rm t}}}%
\def\cd{\ensuremath{c_{\rm d}}}\def\sd{\ensuremath{s_{\rm d}}}%

\begin{document}
\sloppy

\title{Isotropic and anisotropic pointing models}

\author{Andr\'as P\'al \and Kriszti\'an Vida \and L\'aszl\'o M\'esz\'aros \and Gy\"orgy Mez\H{o}}

\institute{%
A. P\'al, K. Vida, L. M\'esz\'aros, Gy. Mez\H o
        \at Konkoly Observatory of the MTA Research Centre for Astronomy and Earth Sciences, Budapest, Hungary \\
E-mail: apal@szofi.net \\
A. P\'al,  L. M\'esz\'aros
        \at Department of Astronomy, E\"otv\"os Lor\'and University, Budapest, Hungary
}

\maketitle

\begin{abstract}
This paper describes an alternative approach for generating 
pointing models for telescopes equipped with serial kinematics, 
esp. equatorial or alt-az mounts. Our model construction does not exploit
any assumption for the underlying physical constraints of the mount,
however, one can assign various effects to the respective
components of the equations. In order to recover the pointing model parameters, 
classical linear least squares fitting procedures can be applied.
This parameterization also lacks any kind of parametric singularity. 
We demonstrate the efficiency of this
type of model on real measurements with meter-class telescopes where
the results provide a root mean square accuracy of $1.5-2$ arcseconds. 
\keywords{Instrumentation: miscellaneous \and Pointing Models \and Methods: analytical}
\end{abstract}

\section{Introduction}
\label{sec:introduction}

Pointing models are widely exploited in telescope control 
in order to correct for mechanical and manufacturing imperfections 
of the various parts of the system. The basic concept of a pointing 
model is to figure out and quantify the difference between the targeted
and the apparent position of the telescope throughout the observations.
This quantification can be done using lookup tables with some grid spacing
as well as using analytical approaches where these functions are related 
to the physical behaviour of the telescope mechanics.

In most of the pointing models, either in the case of equatorial 
mounts \citep{spillar1993}, altitude-azimuth mounts 
\citep{granzer2012,zhang2001} or complex systems
of more telescopes \citep{gothe2013} the raw, uncorrected positions of 
the various axes (hour axis, declination
axis, azimuth axis or elevation axis) are read from high precision
rotary encoders. Such encoders provide a precision within fractions of
arcseconds or even hundredths of arcseconds. There are other types of 
pointing models that involve MEMS accelerometers \citep{meszaros2014} 
instead of such encoders. The attainable precision is much lower in the
case of these integrated accelerometers (currently it is in the range of
an arcminute). Despite the lower precision, these solutions also 
have advantages, including the redundant operation (via the 
various constraints between the accelerometer channels), the 
resistance to unintentional tampering and the possibility of 
installation without modifying or altering the drivers.

The goal of our work presented here is to provide a strictly mathematical 
approach for finding analytical pointing models which provide an 
RMS accuracy within a few arcseconds. 
First, in Sec.~\ref{sec:isotropic} we detail the mathematical 
formalism used in the derivation of an isotropic pointing model. This model
is then extended in Sec.~\ref{sec:anisotropic} with terms that quantify
a generalized form of dependence by performing expansions of spherical
harmonics. These expansions allow the model to take into account various
forms of anisotropic behaviour, for instance the torques inducted by
gravity or misalignments of the gearing. In Sec.~\ref{sec:implementation} we 
report results of test measurements validating these models, as well as 
we detail how the pointing model can be implemented. Finally, our work is
summarized in Sec.~\ref{sec:summary}.

\section{Isotropic pointing model}
\label{sec:isotropic}

Basically, telescope mechanics are treated as a serial robot formed
by two subsequent stages where the two stages are performing
(nearly) orthogonal rotation. Throughout this paper we present the
calculations in first equatorial coordinate system, however, a
similar approach can be involved for alt-az mechanics by simply
replacing the variables appropriately. 
In the case of an equatorial mount, the first shaft (whose bearings are
fixed to the ground) is responsible for the rotation around the hour axis 
while the second, perpendicular stage is responsible for setting the 
declination. In the following, we recall briefly the main ideas behind
the isotropic approach used in \cite{meszaros2014}. 

\begin{figure}
\begin{center}
\resizebox{50mm}{!}{\includegraphics{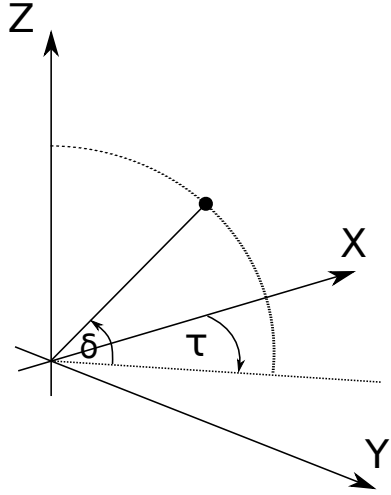}}
\end{center}
\caption{The left-handed coordinate system implied by the 
first equatorial coordinates, i.e. the hour angle $\tau$ and
the declination $\delta$. }
\label{fig:lefthanded}
\end{figure}

Throughout the derivation of our models, let us exploit the 
Cartesian representation of celestial coordinates.
Let us define the standard direction of the telescope tube to the $x+$ 
axis, i.e. without any rotation, the telescope directed to the vector
of $\mathbf{p}_0=(1,0,0)$. This is in accordance with the historic 
parameterization of the first equatorial system if we treat the hour angle 
$\tau$ and declination $\delta$ values as polar angles. However, one
should keep in mind that this parameterization yields a left-handed
$(x,y,z)$ system of reference (see Fig.~\ref{fig:lefthanded}).
An arbitrary point at the sky is then defined by the vector
\begin{equation}
\mathbf{p}=\begin{pmatrix}\cos\delta\cos\tau\\\cos\delta\sin\tau\\\sin\delta\end{pmatrix}\label{eq:pdef}
\end{equation}
while $\mathbf{p}(\tau=0,\delta=0)\equiv\mathbf{p}_0$. 
If one treats the whole telescope system as a series of active rotations, 
then one can write
\begin{equation}
\mathbf{p}(\tau,\delta)=\mathbf{P}_{\rm t}\cdot\mathbf{P}_{\rm d}\cdot\mathbf{p}_0\label{eq:tdnaive}
\end{equation}
where the rotation matrices $\mathbf{P}_{\rm t}$ and $\mathbf{P}_{\rm d}$ are
defined as 
\begin{eqnarray}
\mathbf{P}_{\rm t} & = & \begin{pmatrix}\cos\tau&-\sin\tau&0\\\sin\tau&\cos\tau&0\\0&0&1\end{pmatrix}, \\
\mathbf{P}_{\rm d} & = & \begin{pmatrix}\cos\delta&0&-\sin\delta\\0&1&0\\\sin\delta&0&\cos\delta\end{pmatrix}.
\end{eqnarray}
As it can be recognized, the assumption 
in Eq.~(\ref{eq:tdnaive}) is valid only in ideal circumstances. In real 
applications, the various deflections presented in the telescope
system are quantified by the rotation matrices $\mathbf{H}$, $\mathbf{X}$ and 
$\mathbf{T}$ using the relation
\begin{equation}
\mathbf{p}^\prime(\tau,\delta)=\big(\mathbf{H}\cdot\mathbf{P}_{\rm t}\cdot\mathbf{X}\cdot\mathbf{P}_{\rm d}\cdot\mathbf{T}\big)\cdot\mathbf{p}_0\label{eq:tdmatrices}
\end{equation}
Here,  $\mathbf{H}$ quantifies the misalignments of the telescope hour axis with
respect to the ground, $\mathbf{X}$ quantifies the misalignments of the
declination axis with respect to the hour axis while $\mathbf{T}$ 
quantifies the misalignments of the detector or the tube with respect to
the declination axis. 
These matrices of $\mathbf{H}$, $\mathbf{X}$ and $\mathbf{T}$ are quite
close to unity. Hence, a first-order series expansion can be involved in
order to approximate these. The idea is based on the well-known relation between
skew-symmetric and orthogonal matrices. Namely, for any skew-symmetric
matrix $\mathbf{A}=-\mathbf{A}^{\rm T}$, the exponential of $\exp(\mathbf{A})$
is always an orthogonal matrix with unity determinant. 
In practice, let us write these in the form of
\begin{eqnarray}
\mathbf{H} & = & \exp\begin{pmatrix}0&-c&b\\c&0&-a\\-b&a&0\end{pmatrix}
\approx\begin{pmatrix}1&-c&b\\c&1&-a\\-b&a&1\end{pmatrix}\\
\mathbf{X} & = & \exp\begin{pmatrix}0&-f&e\\f&0&-d\\-e&d&0\end{pmatrix}
\approx\begin{pmatrix}1&-f&e\\f&1&-d\\-e&d&1\end{pmatrix}\\
\mathbf{T} & = & \exp\begin{pmatrix}0&-i&h\\i&0&-g\\-h&g&0\end{pmatrix}
\approx\begin{pmatrix}1&-i&h\\i&1&-g\\-h&g&1\end{pmatrix}
\end{eqnarray}
We call the pointing model \emph{isotropic} 
if the matrices $\mathbf{H}$, $\mathbf{X}$ and $\mathbf{T}$ do 
not depend on the telescope position. 

%% %% %% %% %% %% %% %% %% %% %% %% %% %% %% %% %% %% %% %% %% %% %% %% %% %% 
%% %% %% %% %% %% %% %% %% %% %% %% %% %% %% %% %% %% %% %% %% %% %% %% %% %% 

Here we give the full expansion of $\mathbf{P}$ up to 
the first order. In the formula 
presented below, $\ct$, $\st$, $\cd$ and $\sd$ denote
$\cos\tau$, $\sin\tau$, $\cos\delta$ and $\sin\delta$, respectively:
\begin{eqnarray}
\mathbf{P} & \approx & 
  \begin{pmatrix}\ct\cd&-\st&-\ct\sd\\\st\cd&\ct&-\st\sd\\\sd&0&\cd\end{pmatrix}
+a\begin{pmatrix}0&0&0\\-\sd&0&-\cd\\\st\cd&\ct&-\st\sd\end{pmatrix}\nonumber\\
& + &     b\begin{pmatrix}\sd&0&\cd\\0&0&0\\-\ct\cd&\st&\ct\sd\end{pmatrix}
        +c'\begin{pmatrix}-\st\cd&-\ct&\st\sd\\\ct\cd&-\st&-\ct\sd\\0&0&0\end{pmatrix}\nonumber \\
& + &     d\begin{pmatrix}\st\sd&0&\st\cd\\-\ct\sd&0&-\ct\cd\\0&1&0\end{pmatrix}
        +e'\begin{pmatrix}\ct\sd&0&\ct\cd\\\st\sd&0&\st\cd\\-\cd&0&\sd\end{pmatrix}\nonumber \\
& + &     g\begin{pmatrix}0&-\ct\sd&\st\\0&-\st\sd&-\ct\\0&\cd&0\end{pmatrix}
        +i\begin{pmatrix}-\st&-\ct\cd&0\\\ct&-\st\cd&0\\0&-\sd&0\end{pmatrix}.\label{eq:pexpand}
\end{eqnarray}
During the expansion, the matrices proportional to $c$ and $f$ are going to
be exactly the same like the matrices proportional to $e$ and $h$.
Hence, the above equation depends only on $c+f$ and $e+h$. Therefore,
in the following we use $c'=c+f$ and $e'=e+h$ in our computations. 

If the expansion above (Eq.~\ref{eq:pexpand}) is multiplied by
the reference vector $\mathbf{p}_0$, the term proportional to $g$ will 
also vanish. This term effects computations only when the information
provided by the apparent field rotation is also included in the pointing model
since the first column of the respective matrix is identical to zero.
All in all, the expansion of Eq.~(\ref{eq:tdmatrices}) is 
\begin{eqnarray}
\mathbf{p}^\prime(\tau,\delta) & = & 
\begin{pmatrix}\ct\cd\\\st\cd\\\sd\end{pmatrix}+
a	\begin{pmatrix}0\\-\sd\\\cd\st\end{pmatrix}+ 
b	\begin{pmatrix}\sd\\0\\-\cd\ct\end{pmatrix} \nonumber \\
 & + & c'\begin{pmatrix}-\cd\st\\\cd\ct\\0\end{pmatrix}+
d	\begin{pmatrix}\sd\st\\-\sd\ct\\0\end{pmatrix} \nonumber \\
 & + & e'\begin{pmatrix}\sd\ct\\\sd\st\\-\cd\end{pmatrix}+
i	\begin{pmatrix}-\st\\\ct\\0\end{pmatrix}.\label{eq:tdvectexpansion}
\end{eqnarray}
This equation can be written in the form of
\begin{equation}
\mathbf{p}^\prime(\tau,\delta)=\mathbf{p}(\tau,\delta)+
\sum\limits_{k=1}^{6}a_k\mathbf{p}_k(\tau,\delta),\label{eq:pexpansion}
\end{equation}
where the vector $\mathbf{p}(\tau,\delta)$ is defined by Eq.~(\ref{eq:pdef}),
the parameter vector of $\mathbf{a}=(a_1,\dots,a_6)$ is equivalent to
$(a,b,c',d,e',i)$
and the $\mathbf{p}_k(\tau,\delta)$ are defined accordingly to 
Eq.~(\ref{eq:tdvectexpansion}). 

The parameters ($a_k$) can be obtained by a linear least squares regression
if a series of observations is known with reported mount 
axis encoder positions of $(\tau_n,\delta_n)$
as well as the respective \emph{apparent} positions of $(T_n,D_n)$. Here,
``apparent'' refers to the hour angle and declination values after correcting
for precession, nutation, aberration and refraction as well as
one should accurately obtain the ${\rm DUT1}={\rm UT1}-{\rm UTC}$
differences throughout the derivation of the local sidereal time instances
\citep[see][for more details]{wallace2002,wallace2008}.
Once this series of $1\le n \le N$ measurements is known, one has to minimize
the merit function
\begin{equation}
\chi^2=\sum\limits_{n=1}^{N}\chi^2_n\label{eq:chi2sum}
\end{equation}
where
\begin{equation}
\chi^2_n=\left[\mathbf{p}(\tau_n,\delta_n)+\sum\limits_{k=1}^{6}a_k\mathbf{p}_k(\tau_n,\delta_n)-\mathbf{p}(T_n,D_n)\right]^2.\nonumber
\end{equation}
Minimization of the above equation for $\chi^2$ yields a linear array
of equations whose solution is then straightforward. Due to the three dimensional
nature of the $\mathbf{p}$ vectors, this equation for $\chi^2$ implies $3N$
independent terms. However, the effective number for degrees of freedom
is only $2N-6$ since for all $1\le k\le 6$ values and for all observations,
\begin{equation}
\mathbf{p}(\tau_n,\delta_n)\cdot\mathbf{p}_k(\tau_n,\delta_n)=0.
\end{equation}
In other words, these 6 base vectors of $\mathbf{p}_k$ defined in the 
expansion (\ref{eq:pexpansion}) are always perpendicular to the
observation direction $\mathbf{p}(\tau_n,\delta_n)$.

In the following we extend the isotropic model described above with
terms that provide significantly better accuracy.

%% %% %% %% %% %% %% %% %% %% %% %% %% %% %% %% %% %% %% %% %% %% %% %% %% %% 
\begin{table}
\caption{Coefficients of the expansion of the pointing model
up to the first order in the spherical harmonics. 
Note that the terms marked with stars ($\bigstar$)
and dots ($\bullet$) are not linearly independent. Hence, the
first order expansion of the pointing model has 22 independent
coefficients instead of 24.}
\label{table:ypointingcoeffs}
\begin{center}\begin{tabular}{l|llll}
	& $Y_{00}$ & $Y_{1,-1}$ & $Y_{1,0}$ & $Y_{0,1}$ \\[5pt]
\hline
$a$	& $\begin{pmatrix}0\\-\sd\\\cd\st\end{pmatrix}$ 
	& $\begin{pmatrix}0\\-\sd\cd\st\\\cd^2\st^2\end{pmatrix}$%\hspace*{-2mm}
	& $\begin{pmatrix}0\\-\sd^2\\\sd\cd\st\end{pmatrix}$
	& $\begin{pmatrix}0\\-\sd\cd\ct\\\cd^2\st\ct\end{pmatrix}_{\bullet}$\\[18pt]%\hspace*{-3mm}
$b$	& $\begin{pmatrix}\sd\\0\\-\cd\ct\end{pmatrix}$%\hspace*{-5mm}
	& $\begin{pmatrix}\sd\cd\st\\0\\-\cd^2\st\ct\end{pmatrix}_{\bullet}$%\hspace*{-6mm}
	& $\begin{pmatrix}\sd^2\\0\\-\sd\cd\ct\end{pmatrix}$%\hspace*{-7mm}
	& $\begin{pmatrix}\sd\cd\ct\\0\\-\cd^2\ct^2\end{pmatrix}$ \\[18pt]
$c'$	& $\begin{pmatrix}-\cd\st\\\cd\ct\\0\end{pmatrix}$%\hspace*{-5mm}
	& $\begin{pmatrix}-\cd^2\st^2\\\cd^2\st\ct\\0\end{pmatrix}$ 
	& $\begin{pmatrix}-\sd\cd\st\\\sd\cd\ct\\0\end{pmatrix}_{\bullet}$%\hspace*{-7mm}
	& $\begin{pmatrix}-\cd^2\st\ct\\\cd^2\ct^2\\0\end{pmatrix}$ \\[18pt]
$d$	& $\begin{pmatrix}\sd\st\\-\sd\ct\\0\end{pmatrix}_{\bigstar}$%\hspace*{-5mm}
	& $\begin{pmatrix}\sd\cd\st^2\\-\sd\cd\st\ct\\0\end{pmatrix}$%\hspace*{-5mm}
	& $\begin{pmatrix}\sd^2\st\\-\sd^2\ct\\0\end{pmatrix}$%\hspace*{-5mm}
	& $\begin{pmatrix}\sd\cd\st\ct\\-\sd\cd\ct^2\\0\end{pmatrix}$ \\[18pt]
$e'$	& $\begin{pmatrix}\sd\ct\\\sd\st\\-\cd\end{pmatrix}$ 
	& $\begin{pmatrix}\sd\cd\st\ct\\\sd\cd\st^2\\-\cd^2\st\end{pmatrix}$%\hspace*{-2mm}
	& $\begin{pmatrix}\sd^2\ct\\\sd^2\st\\-\sd\cd\end{pmatrix}$%\hspace*{-5mm}
	& $\begin{pmatrix}\sd\cd\ct^2\\\sd\cd\st\ct\\-\cd^2\ct\end{pmatrix}$ \\[18pt]
$i$	& $\begin{pmatrix}-\st\\\ct\\0\end{pmatrix}$ 
	& $\begin{pmatrix}-\cd\st^2\\\cd\st\ct\\0\end{pmatrix}$ 
	& $\begin{pmatrix}-\sd\st\\\sd\ct\\0\end{pmatrix}_{\bigstar}$%\hspace*{-5mm}
	& $\begin{pmatrix}-\cd\st\ct\\\cd\ct^2\\0\end{pmatrix}$ 
\end{tabular}\end{center}
\end{table}
%% %% %% %% %% %% %% %% %% %% %% %% %% %% %% %% %% %% %% %% %% %% %% %% %% %% 

\section{Anisotropic extensions}
\label{sec:anisotropic}

The idea behind the isotropic pointing model is the fact that the
parameters $a_k$ of the respective expansion do not depend on the 
the pointing direction defined by $\tau$ and/or $\delta$. In order
to attain better accuracy, we investigate now how 
the matrices $\mathbf{H}$, $\mathbf{X}$ and $\mathbf{T}$ and hence 
the parameters $a_k$ can depend on the position of the telescope. 
Recalling earlier works \citep[see e.g.][]{zhang2001},
we examine the expansion of these terms via spherical harmonics.
Namely, each of the $a_k$ constants are replaced by a linear combination
of spherical harmonics up to a certain order $L$. Here $L$ is the maximum
order of the corresponding spherical harmonics and the respective
Legendre polynomials (see also later on for the actual definitions). 
In practice, let us extend Eq.~(\ref{eq:pexpansion}) as
\begin{equation}
\mathbf{p}^\prime(\tau,\delta) = \mathbf{p}(\tau,\delta)+ 
\sum\limits_{k=1}^{6}\sum\limits_{\ell=0}^{L}\sum\limits_{m=-\ell}^{\ell}
a_{k\ell m}Y_{\ell m}(\tau,\delta)\mathbf{p}_k(\tau,\delta), \label{eq:pyyexpansion}
\end{equation}
where $Y_{\ell m}(\tau,\delta)$ represent the real orthogonal 
spherical harmonics with the indices $(\ell,m)$:
\begin{equation}
Y_{\ell m}(\tau,\delta)=
\begin{cases}
  K_{\ell m}P_\ell^{-m}(\sin\delta) \sin(m\tau) &\mbox{if } m<0 \\
  K_{\ell m}P_\ell^m(\sin\delta) \cos(m\tau) & \mbox{if } 0\ge m.
\end{cases}
\end{equation}
In the above definition, the terms $P_\ell^m(\cdot)$ denote the 
associated Legendre polynomials.
The constants $K_{\ell m}$ can have arbitrary but non-zero values 
since in Eq.~(\ref{eq:pyyexpansion}), the $Y_{\ell m}$ functions
are multiplied by an unknown parameter. The series for the first
three orders are
\begin{eqnarray}
Y_{00} & = & 1, \\
Y_{1,-1} & = & \cos\delta\sin\tau, \\
Y_{10} & = & \sin\delta, \\
Y_{1,1} & = & \cos\delta\cos\tau, \\
Y_{2,-2} & = & \cos^2\delta\sin(2\tau), \\
Y_{2,-1} & = & \cos\delta\sin\delta\sin\tau, \\
Y_{20} & = & 3\sin^2\delta-1, \\
Y_{21} & = & \cos\delta\sin\delta\cos\tau, \\
Y_{22} & = & \cos^2\delta\cos(2\tau). 
\end{eqnarray}
In Table~\ref{table:ypointingcoeffs}, we also summarize the terms
$Y_{\ell m}(\tau,\delta)\mathbf{p}_k(\tau,\delta)$ up to the order
of $\ell \le L=1$. 

\subsection{Parameter ambiguities}
\label{sec:ambiguities}

It is essential to note that an important property of the terms 
$Y_{\ell m}(\tau,\delta)\mathbf{p}_k(\tau,\delta)$ is the ambiguity
due to the lack of linear independence between some of the terms. 
For instance, up to the order of $L=1$, there are two relations that
should be taken into account:
\begin{eqnarray}
0 & = & \mathbf{p}_4 + \mathbf{p}_6Y_{10}, \label{eq:l1lin1}\\
0 & = & \mathbf{p}_2Y_{1,-1} + \mathbf{p}_3Y_{10} + \mathbf{p}_1Y_{11}\label{eq:l1lin2}
\end{eqnarray}
These relations are valid for arbitrary values of $(\tau,\delta)$ and hence
in the first case (Eq.~\ref{eq:l1lin1}) either $a_{400}$ or $a_{610}$ 
should be omitted from the model fit. In the second case (Eq.~\ref{eq:l1lin2}),
one of the terms $a_{21,-1}$, $a_{310}$ or $a_{111}$ should be omitted.
We also mark these terms affected by the linear dependence 
in Table~\ref{table:ypointingcoeffs}. Due to the presence of these two
equations, the $L=1$ anisotropic pointing model 
has $6\cdot4-2=22$ free parameters in total. 

For completeness, we give here the six 
additional respective relations which appear if we consider spherical
harmonics expansion up to the order of $L=2$:
\begin{eqnarray}
0 & = & \mathbf{p}_6Y_{2,-1} + \mathbf{p}_4Y_{1,-1}, \label{eq:yl2r1}\\
0 & = & \mathbf{p}_6Y_{21} + \mathbf{p}_4Y_{11}, \\
0 & = & \mathbf{p}_6Y_{20}+\mathbf{p}_6+3\mathbf{p}_4Y_{10}, \\
0 & = & 3\mathbf{p}_1Y_{21}+\mathbf{p}_3+3\mathbf{p}_2Y_{2,-1}+\mathbf{p}_3Y_{20}, \\
0 & = & 3\mathbf{p}_1Y_{22}-6\mathbf{p}_1+3\mathbf{p}_2Y_{2,-2}- \nonumber \\
  & & -\mathbf{p}_1Y_{20}+6\mathbf{p}_3Y_{21}, \\
0 & = & 3\mathbf{p}_2Y_{22}-6\mathbf{p}_2+3\mathbf{p}_1Y_{2,-2}- \nonumber \\
  & & -\mathbf{p}_2Y_{20}+6\mathbf{p}_3Y_{2,-1}\label{eq:yl2r6}
\end{eqnarray}
The omission of the respective $a_{k\ell m}$ parameters should be done according
to these relations. For example, the first respective parameters in 
Eqs.~(\ref{eq:yl2r1})--(\ref{eq:yl2r6}), i.e. $a_{62,-1}$, $a_{621}$,
$a_{620}$, $a_{121}$, $a_{a122}$ and $a_{222}$ can forcibly be set to zero
during the fitting procedures (along with, for instance, the parameters $a_{610}$ and $a_{111}$,
see above). Therefore, the $L=2$ anisotropic pointing model 
has $9\cdot4-2-6=46$ free parameters in total.

\section{Interpretation of the model coefficients}
\label{sec:interpretation}

Our initial attempt was to create a pointing model that is completely
based on a direct mathematical approach. However, the various quantities
appearing in the expansion, i.e. the coefficients of 
$a_{k\ell m}$ as well as the respective functions of 
$Y_{\ell m}(\tau,\delta)\mathbf{p}_k(\tau,\delta)$ can be assigned
to various physical characteristics of the telescope mount. 
In the following, we briefly summarize these ``roles'' of the various 
$a_{k\ell m}$ coefficients. 

\subsection{The isotropic case}
\label{sec:interpretation:isotropic}

As it is known from earlier works \citep[see e.g.][]{spillar1993,buie2003}, 
the most essential parameters of a telescope pointing correction functions
are related to the polar displacement, the encoder zero point offsets and
the non-orthogonality deviation of the two axes.
Indeed, it can be recognized that $a\equiv a_{1}\equiv a_{100}$ 
and $b\equiv a_{2}\equiv a_{200}$ are the terms that are proportional to the 
polar displacement (in azimuthal and elevation directions, respectively),
$c'\equiv a_{3}\equiv a_{300}$ is the hour angle encoder offset, 
$d\equiv a_{4}\equiv a_{400}$ is the deviation from the right angle
between the two axes and $e\equiv a_{5}\equiv a_{500}$ is the declination
encoder offset. The term $i\equiv a_{6}\equiv a_{600}$ is the deviation
of the optical axis in the direction of sidereal rotation. 

\subsection{Anisotropic models}
\label{sec:interpretation:anisotropic}

One of the most frequently cited anisotropic properties of a telescope
mount arises from the gravity of Earth. Gravity yields a deflection of
the telescope tube directly as well as indirectly via the unbalanced axes. 
As the barycenter of the various parts moves along 
as the telescope is slewing to various celestial positions, the net torque
also results in anisotropic pointing offsets. 
In practice, the equatorial and declination stages are usually 
driven by worm gears. Hence, another kind of anisotropy appears in the 
mechanical system if we consider the ellipticity and/or eccentricity 
of the driven worm gears of these axes. Further details and examples
about the interpretation of various anisotropic coefficients 
are found below, in Sec.~\ref{sec:tests}.

\subsection{Periodic errors}
\label{sec:interpretation:periodic}

An expansion by spherical harmonics can be used to include additional
terms that have a dedicated physical relevance. A prominent example can be
the characterization of periodic errors. Most of the classic telescope mounts
are worm-geared systems and due to the limited resolution of the rotary
encoders, raw hour angle and declination values are recovered from 
multi-turn encoders mounted to the worm shaft instead of the gear shaft. 
Let us suppose a gear ratio of $G$. As the term $a_3\mathbf{p}_3$ quantifies
the encoder zero point offset in the hour angle, it can be 
deduced that the two additional terms of 
\begin{equation}
a_{3,G,-G}\mathbf{p}_3Y_{G,-G}+a_{3,G,G}\mathbf{p}_3Y_{G,G}
\end{equation}
characterize the periodic error. Namely, 
\begin{equation}
\sqrt{a_{3,G,-G}^2+a_{3,G,G}^2}
\end{equation}
is proportional to the amplitude of the periodic error while the angle
\begin{equation}
\mathrm{arg}(a_{3,G,G},a_{3,G,-G})
\end{equation}
tells us the phase of the periodic error at $\tau=0$ encoder position. 

%% %% %% %% %% %% %% %% %% %% %% %% %% %% %% %% %% %% %% %% %% %% %% %% %% %% 
\begin{figure}
\begin{center}
\resizebox{105mm}{!}{\includegraphics{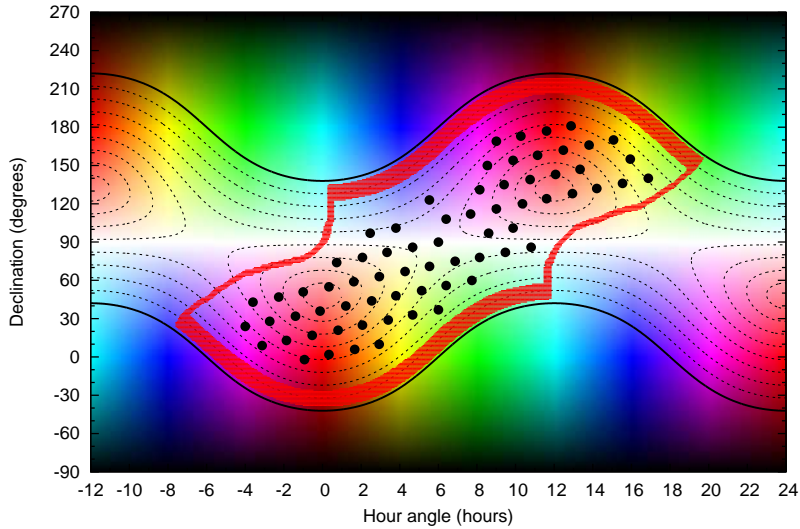}}
\end{center}
\caption{The sky in the first equatorial system,
as it is seen by the RCC telescope mount used for the evaluation
of the pointing models. The hue levels
(in parallel with the horizontal axis) represent the celestial hour angle
values while the luminance levels correspond to the celestial declination values.
The distance in the color space is equivalent to the apparent celestial
distance while the projected distance in this graph is proportional to
the movements of the equatorial and declination axes. The clear white
stripe is the distorted projection of the north celestial pole while the fully
black stripe (at the bottom and the top) is the distorted projection
of the south celestial pole. The horizon is shown by the two parallel
thick black curves while the various positive horizontal elevation
levels (in steps of $10^\circ$)
are shown as thin black dashed curves. The mount domain is limited by
the thick red stripe. These limits correspond both to the horizon
as well as the northern pillar of the equatorial axis pier.}
\label{fig:rccaxes}
\end{figure}
%% %% %% %% %% %% %% %% %% %% %% %% %% %% %% %% %% %% %% %% %% %% %% %% %% %% 

\section{Implementation and model tests}
\label{sec:implementation}

In order to use the pointing model in practice, one should
consider how it can be implemented for ``real'' telescopes. 
The actual implementation has to have two essential parts: 
\begin{itemize}
\item First, one should obtain an appropriately chosen series of 
astrometric calibration frames of the given instrument in order to 
obtain the respective $(\tau_n,\delta_n)$ and $(T_n,D_n)$ values
needed for the model. Afterwards, once these data are known, the
parameters of the pointing model are needed to be fitted.
\item Second, ``implementation'' also refers to the 
insertion of the model parameters into the telescope control system (TCS)
in order to have a benefit during the observations.
\end{itemize}
In this section, we summarize our method of 
implementation, considering both the model parameter regression as
well as the integration in a TCS.

\subsection{Derivation of pointing model parameters}
\label{sec:modelfit}

To implement our model, one should
employ a classic, purely linear least squares regression analysis on 
functions containing merely linear combinations of trigonometric expressions. 
By considering the definitions of the linear least squares method 
($\chi^2$, see Eq.~\ref{eq:chi2sum}) as well as the actual model 
(see Eq.~\ref{eq:pyyexpansion}), one could derive that the
merit function needed to be minimized is 
\begin{equation}
\chi^2=\sum\limits_{n=1}^N \left[\mathbf{p}_{L=\dots}(\tau_n,\delta_n)
-\mathbf{p}(T_n,D_n)
\right]^2,
\end{equation}
where
\begin{equation}
\mathbf{p}_{L=\dots}(\tau,\delta) = \mathbf{p}(\tau_n,\delta_n)+
\sum\limits_{k=1}^{6}\sum\limits_{\ell=0}^{L}\sum\limits_{m=-\ell}^{\ell}
a_{k\ell m}Y_{\ell m}(\tau,\delta)\mathbf{p}_k(\tau,\delta).
\end{equation}
If $L=0$, we get the isotropic pointing model (having $6$ parameters
in total) while for $L=1$ and $L=2$, we get the anisotropic cases with 
$22$ and $46$ free parameters. Of course, the value of $L$ can further be
increased, however, one should keep in mind that additional 
identities similar to Eqs.~(\ref{eq:l1lin1})-(\ref{eq:l1lin2}) and/or 
Eqs.~(\ref{eq:yl2r1})-(\ref{eq:yl2r6}) will appear during the expansions. 

Our current implementation for minimizing $\chi^2$ is based on 
shell scripts written in the language of \texttt{bash}. These scripts
exploit standard UNIX text processing utilities as well as the 
\texttt{lfit} utility of the FITSH package \citep{pal2012}. 
This implementation expects a four-column input file (where each line 
contains the $\tau_n,\delta_n,T_n,D_n$ values) and it is available 
on the FITSH website\footnote{\url{http://fitsh.szofi.net/}} in the 
``Examples'' section\footnote{\url{http://fitsh.szofi.net/example/pointing/}}.

\subsection{Integration in telescope control systems}
\label{sec:tcsintegration}

Once the pointing model parameters are known, they
should be integrated in the TCS. However, such 
an integration needs a lower level access to the TCS
software stack while the above described procedure for obtaining the 
coefficients can be done by anyone who has observational access to the 
telescope. 

There are two ways to use the model:
\begin{itemize}
\item First, Eq.~(\ref{eq:pyyexpansion}) (or Eq.~\ref{eq:pexpansion} 
when considering only anisotropic models) should be evaluated when
one computes the \emph{apparent} equatorial coordinates based on
the encoder values. This type of calculation is performed when the 
TCS is asked to report the current position, i.e. \emph{after} the
\emph{reading} of the encoders on the mount axes.
\item Second, the \emph{inverse} form of Eq.~(\ref{eq:pyyexpansion})
needs to be evaluated during pointing and tracking of the telescopes.
Namely, if one has to determine the encoder values based 
on the apparent first equatorial coordinates. This type of calculation
is performed when the TCS is commanded to slew the telescope to
a certain position, i.e. \emph{before} targeting the motors (and/or
any mechanisms). 
\end{itemize}
Any software library that handles pointing model evaluation needs to be 
ready for implementing both of the above types. 

In practice, the inversion of Eq.~(\ref{eq:pyyexpansion}) can be implemented by 
changing the sign in the series expansion coefficients. This type of
inversion yields systematic offsets in the order of $a_{k\ell m}^2$. If
the typical mount deflections are in the range of a few arcminutes
(i.e. a milliradian), then the error caused by this improper inversion
yields a systematic effect in the range of sub-arcsecond scale (i.e. 
microradians) which is usually sufficient. If higher accuracy is needed,
the inversion can be done in a two-step iteration, yielding errors
in the range of $a_{k\ell m}^3$. Since practical applications do not
rely on subsequent evaluation of the pointing model corrections, these
errors do not accumulate.

The above cited ``Examples'' section of the FITSH webpage contains a 
full implementation of an ANSI C library that can also be integrated in a TCS. 
In addition, this section of the webpage links some code snippets taken
from the currently running version of an 1-m class telescope control
software (see below the next subsection for more details).

%% %% %% %% %% %% %% %% %% %% %% %% %% %% %% %% %% %% %% %% %% %% %% %% %% %% 
\begin{figure}
\begin{center}
\resizebox{57.5mm}{!}{\includegraphics{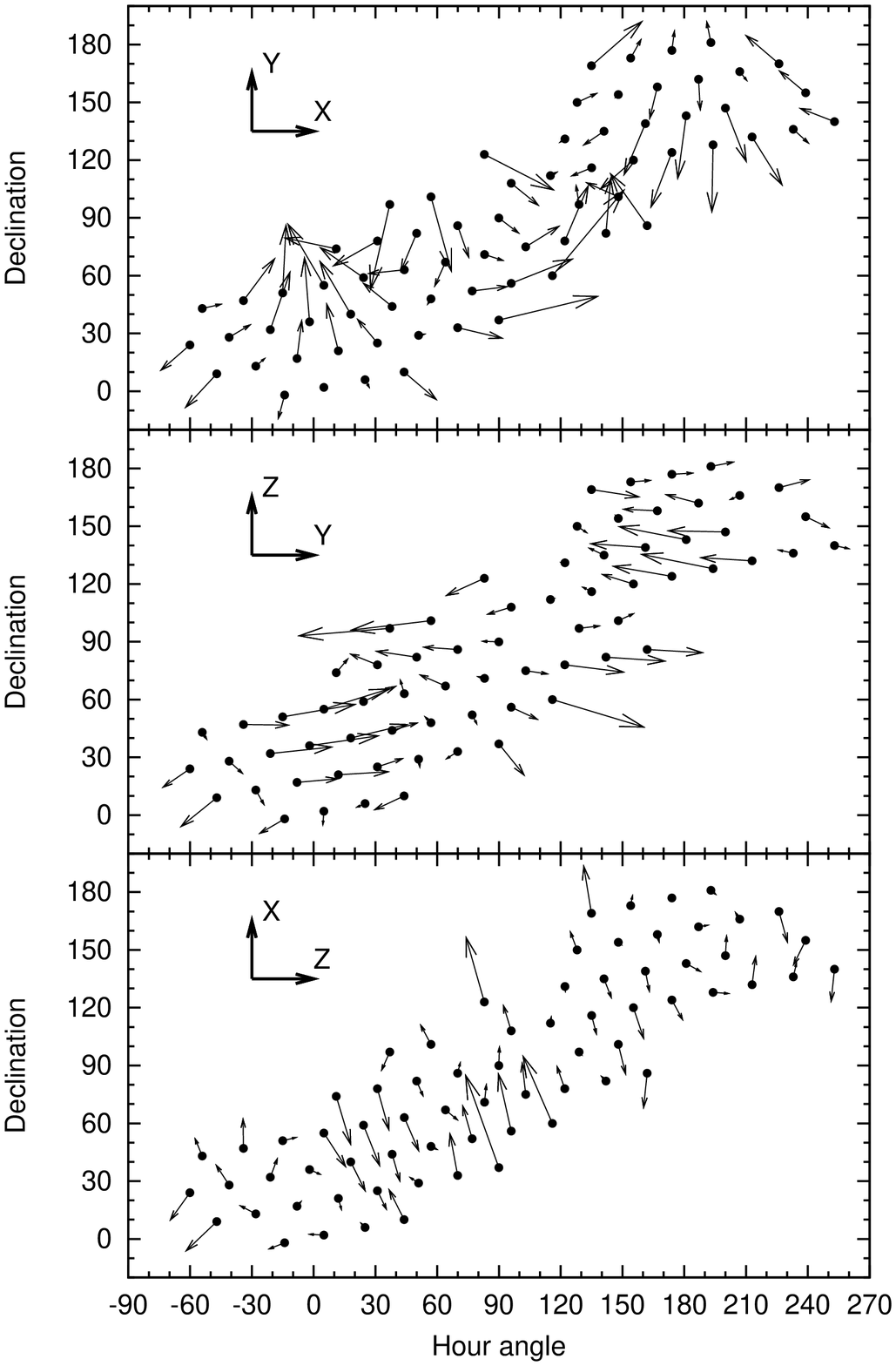}}%
\resizebox{57.5mm}{!}{\includegraphics{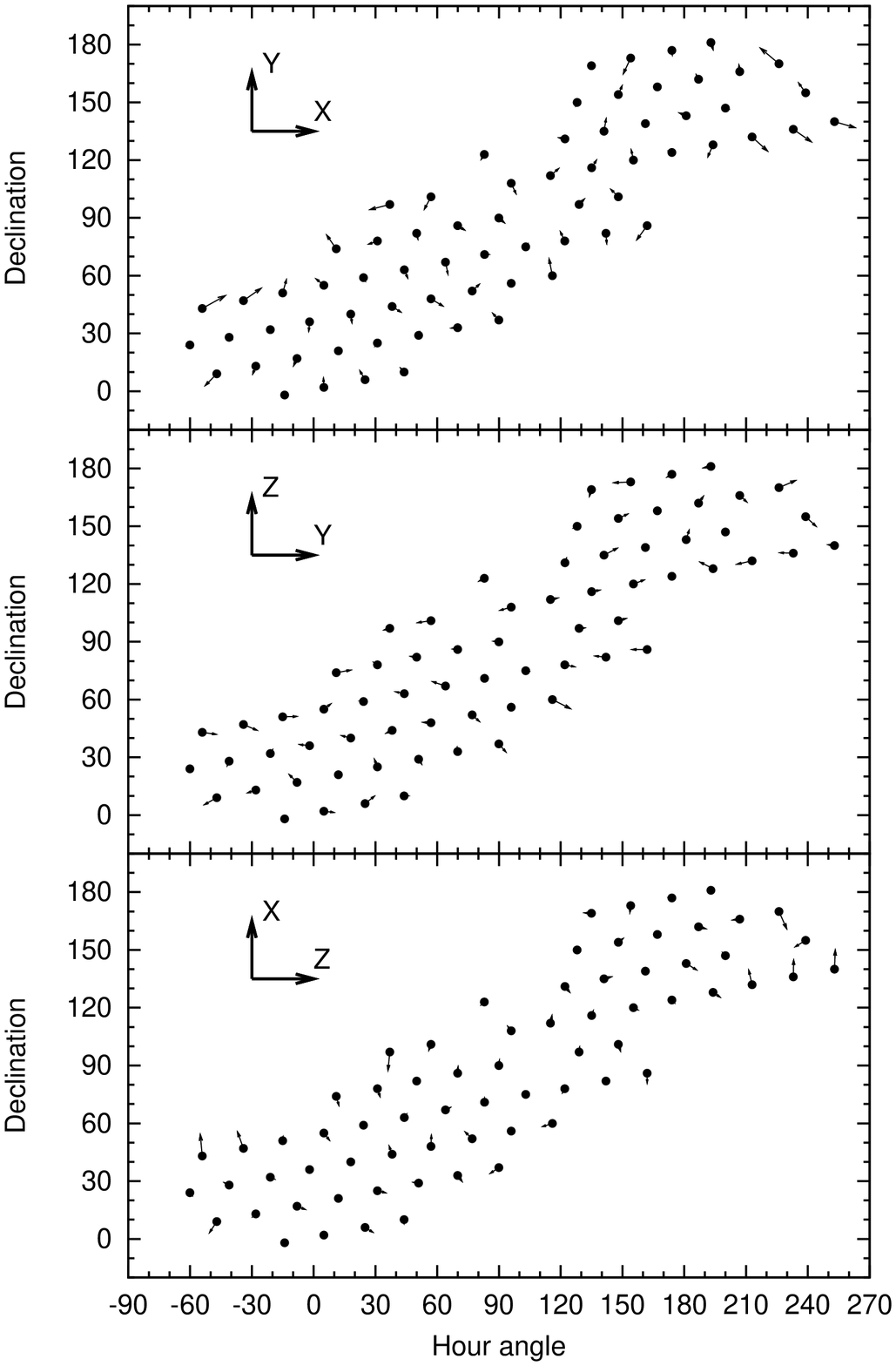}}
\end{center}
\caption{Residuals of the pointing model applied to the Konkoly 1-meter
RCC telescope. The left panel shows the respective components of the
$\mathbf{p}_{L=0}(\tau_n,\delta_n)-\mathbf{p}(T_n,D_n)$ vectors for 
the isotropic case while the right panel shows these difference vector 
components for the $L=1$ anisotropic model. The scales for the difference
vectors are the same for both panels: one degree in the main 
abscissa and ordinate refers to one arcsecond in the residual vector
components. Note that the figures are redundant, not only because they show
all of the $(x,y)$, $(y,z)$ and $(z,x)$ component combinations but 
due to the fact that residual vectors are always tangential to the 
unit sphere. The strong structure of systematics in the residual in 
the isotropic case is quite explicit. Note also that the residual vectors
would be as large as the individual panels if we plot the residuals 
\emph{without} any pointing model. See text for further details. }
\label{fig:residual}
\end{figure}
%% %% %% %% %% %% %% %% %% %% %% %% %% %% %% %% %% %% %% %% %% %% %% %% %% %% 

\subsection{Model tests}
\label{sec:tests}

In order to test the feasibility of our model we conducted a series 
of observations with the recently refurbished 1-meter 
Ritchey--Chr\'etien-coud\'e (RCC) telescope of the Konkoly Observatory. 
This telescope is installed on an English cross-axis mount. The
northern pillar of this type of mount strictly limits the slew 
domain of the telescope within a certain range
since the tube is unable to observe objects close to lower culmination. 
In the available telescope motion domain we defined a grid of 
($\tau_n,\delta_n$) values and exposures were taken on each grid 
point with a net time of 20 seconds. 
In total, we gathered 81 images covering the sky nearly uniformly 
with the inclusion of polar crossing.
Due to the partially cloudy weather, only 69 of these images have 
successfully been analyzed in order to have accurate astrometric solutions.
The positions of these measurements are displayed in Fig.~\ref{fig:rccaxes}.
Images were acquired with an Andor iXon-888 frame-transfer electron-multiplying
CCD camera (EMCCD) and reduced with the standard calibration procedures
using the FITSH utilities \citep{pal2012}. Since the net field-of-view of
the optical setup is relatively small ($3.4^\prime\times3.4^\prime$), 
astrometric solutions were derived using images acquired in parallel 
with a guider telescope. For an initial solution, we exploited 
the offline version of the Astrometry.net package \citep{lang2010} while
the cross-matching of the two camera images as well as the final astrometric
solutions were obtained with the appropriate tasks of the FITSH package
using the USNO catalogue \citep{monet2003} as a reference. 

After obtaining the J2000.0 centroids of each frame, apparent first
equatorial coordinates (hour angle and declination) were obtained after
correcting for precession, nutation, aberration and refraction using
the standard procedures \citep{meeus1998,wallace2008}. 
The list of encoder positions
read at the exposure midpoint was used as an input series of $(\tau_n,\delta_n)$
values while the corrected astrometric solutions were used as an
input series of $(T_n,D_n)$ values.

The unbiased residuals of the isotropic, the first-order anisotropic 
and the second-order anisotropic pointing models were 
$\sigma_0=69.1\cdot10^{-6}$, $\sigma_1=14.8\cdot10^{-6}$ and
$\sigma_2=10.2\cdot10^{-6}$ radians, respectively (where $\sigma_L$ denotes
the residuals corresponding to the $L$th order in the spherical 
harmonics expansion). Since one arcsecond is equal to
$4.84814\cdot10^{-6}$ radians, these values correspond to $14.2^{\prime\prime}$,
$3.0^{\prime\prime}$ and $2.1^{\prime\prime}$. We note here that
without any pointing model applied (which is similar to setting forcibly all 
of the coefficients $a_{k\ell m}$ to zero), the residual 
is $\sigma_{\emptyset}=0.0011$ radian, equivalent to $3.7^{\prime}$. 
This is $\approx 16$ times larger even than
the residual corresponding to the isotropic ($L=0$) case. 
We plot the typical structures of pointing model residuals for 
the $L=0$ and $L=1$ case in the two panels of Fig.~\ref{fig:residual}.

The fitted values and the respective uncertainties of the model coefficients 
can also be used to measure the actual deflections of the telescope. For
instance, in our case the isotropic model yields values for the polar 
displacement and encoder zero point offsets that can be used for tuning 
the telescope. The actual values for the polar displacement 
are $a=0.000163\pm0.000012$, $b=0.000356\pm0.000010$ while the encoder 
zero points are $c'=0.000844\pm0.000013$ and $e'=0.000705\pm0.000009$ radians. 
It also turns out that the deviation of the two axes is also significant,
being in the range of arcminute ($d=0.000315\pm0.000085$). 
However, this value represents an \emph{average} value for these deflections
since the effects described in Sec.~\ref{sec:interpretation:anisotropic}
can yield an anisotropy in any of the previously mentioned values. Indeed,
for instance, the deviation of the two axes shows a clear dependence on the
telescope position itself, namely the corresponding coefficients are 
$a_{400}=0.000573\pm0.000091$,
$a_{410}=-0.000008\pm0.000023$, 
$a_{41,-1}=-0.000077\pm0.000056$ and
$a_{41,+1}=0.000191\pm0.000056$ (also in radians). It means that the 
deflection of the two
axes can \emph{vary} with a full amplitude of at least $1.5$ arcminutes on 
the total telescope motion domain.

We also included two terms corresponding to the $G=240$ gear ratio of the
hour axis worm-drive of the RCC telescope. This inclusion yielded a 
decrement of the unbiased $\chi^2$ values of $3.5$ (which should be compared to
$2N-P=116$, where $N=69$, the number of the frames and $P=22$, the number
of parameters in the $L=1$ anisotropic model). At the first
glance, this fact might prove that the pointing model is improved by
this additional term. However, compared to 
the total degrees of freedom, this decrement is marginal as well as
the respective amplitude that differs from zero only by 1.5-sigma. In addition,
total measured periodic error of this telescope is $\lesssim0.15^{\prime\prime}$,
which is well below the residuals of the model. All in all, we can 
conclude that \emph{this} telescope cannot benefit with this inclusion,
but other telescopes (e.g. amateur-class mechanics with typical periodic errors
in the range of tens of arcseconds) might do so. 

\subsection{Polar crossing}
\label{sec:polarcrossing}

It should be noted that spherical harmonics yield the same values
while we observe the same celestial position before and after polar crossing.
In other words, $Y_{\ell m}(\tau,\delta)=Y_{\ell m}(\tau+180^\circ,180^\circ-\delta)$
for all values of $\tau$ and $\delta$. However, some anisotropic physical
effects (see Sec.~\ref{sec:interpretation:anisotropic} above) might have
different yields depending on the pier side of the telescope. 
Since the telescope can reach most of the target positions in both 
configurations, it is worth separating the coefficients corresponding to 
these two configuration domains. Here we define two target points to be
in the same configuration if the telescope can be re-positioned between
the two points without polar crossing. 

The easiest way to do this is to consider points with $\delta\lesssim90^\circ$
and $90^\circ\lesssim\delta$ in independent fits. Indeed, our analysis
yields an unbiased residual of $1.6^{\prime\prime}$ and
$1.7^{\prime\prime}$ for the $L=1$ model if the two parts of the sky
are treated separately. Both residuals are significantly smaller than
the residual of $3.0^{\prime\prime}$ and even a bit smaller than the 
$L=2$ model for the full coverage. If we perform this separation in the
isotropic model, the unbiased residuals are going to be roughly the 
same (namely, $14.0^{\prime\prime}$ and $12.1^{\prime\prime}$, in our case).

\section{Summary}
\label{sec:summary}

In this paper we have presented a generic analytical anisotropic telescope
pointing model derived from purely mathematical considerations. Our results
show that this type of approach provides RMS residuals 
comparable and even better than earlier types of pointing models.
It should be noted, however, that this pointing model does not account directly
the temperature dependence of the telescope system. Based on earlier
works \citep{mittag2008}, the coefficients for the drifts of the various 
parameters due to the thermal expansion can be as large as 
$0.2 - 1^{\prime\prime}$ per degrees Celsius. 
Hence, the accuracy of the ambient temperature
should be within a few degrees Celsius in order to recover the 
pointing with an accuracy comparable to the RMS residuals.
Uncertainties in the refraction model, for instance a variation 
of $15\,{\rm hPa}$ in the pressure and/or $5$ degrees Celsius in the 
temperature yield an offset of $1^{\prime\prime}-2^{\prime\prime}$ even at 
moderate ($25-45$ degrees of) elevations above the horizon. Although
an improper refraction model can be corrected via the $Y_{\ell m}$ functions,
the correction factors depend on the temperature and/or the density profile
of the layers in the atmosphere.  
A major advantage of our model is the lack of parameterization via polar 
coordinates. In our tests, one of the grid points had a declination 
of $\delta_{\rm J2000}=89^\circ55^{\prime}$ and it is also a well behaved 
point, not an outlier -- despite the fact that the respective value 
of $1/\cos\delta$ is more than $700$. 
In order to have a successful implementation for our model, one should
employ only classic linear least squares regression analysis on 
functions containing purely trigonometric expressions. The implementation
of both the model fit and the model evaluation functions
is straightforward, and is demonstrated on the FITSH webpage. 

\begin{acknowledgement}
We thank the detailed review and the valuable comments of the anonymous
referee. We also thank Emese Plachy and L\'aszl\'o Moln\'ar the careful 
proof corrections. Our project has been supported by the Hungarian Academy 
of Sciences via the grant LP2012-31 as well as via the Hungarian OTKA grants
K-104607, K-109276 and K-113117.
\end{acknowledgement}

{}

\end{document}